\documentclass[journal,10pt]{IEEEtran}

\IEEEoverridecommandlockouts 
\usepackage{epsfig,latexsym}
\usepackage{subfigure}
\usepackage{float}
\usepackage{indentfirst}
\usepackage{amsmath}
\usepackage{amssymb}
\usepackage{cite}
\usepackage{bm}
\usepackage{url}
\usepackage{lastpage}
\usepackage{color}
\usepackage{fancyhdr}
\usepackage{algorithm}
\usepackage{algorithmic}
\usepackage{multirow}
\usepackage[square, comma, sort&compress, numbers]{natbib} 
\normalsize

\begin{document}

\title{Deep Learning-Aided Dynamic Read Thresholds Design For Multi-Level-Cell Flash Memories}


\author{Zhen Mei, Kui Cai,~\IEEEmembership{Senior Member,~IEEE} and Xuan He,~\IEEEmembership{Member,~IEEE}
\thanks{The material
in this paper was presented in part at IEEE International Conference on Communications (ICC), May 2019. Zhen Mei, Kui Cai and Xuan He are with the Science and Math Cluster, Singapore University of Technology and Design, Singapore 487372 (email: mei\_zhen@outlook.com; cai\_kui@sutd.edu.sg; xuan\_he@sutd.edu.sg ).} 

} 

\maketitle

\begin{abstract}
The practical NAND flash memory suffers from various non-stationary noises that are difficult to be predicted. Furthermore, the data retention noise induced channel offset is unknown during the readback process. This severely affects the data recovery from the memory cell. In this paper, we first propose a novel recurrent neural network (RNN)-based detector to effectively detect the data symbols stored in the multi-level-cell (MLC) flash memory without any prior knowledge of the channel. However, compared with the conventional threshold detector, the proposed RNN detector introduces much longer read latency and more power consumption. To tackle this problem, we further propose an RNN-aided
(RNNA) dynamic threshold detector, whose detection thresholds can be derived based on the outputs of the RNN detector. We thus only need to activate the RNN detector periodically when the system is idle.
 Moreover, to enable soft-decision decoding of error-correction codes, we first show how to obtain more read thresholds based on the hard-decision read thresholds derived from the RNN detector. We then propose integer-based reliability mappings based on the designed read thresholds, which can generate the soft information of the channel. Finally, we propose to apply density evolution (DE) combined with differential evolution algorithm to optimize the read thresholds for LDPC coded flash memory channels. Computer simulation results demonstrate the effectiveness of our RNNA dynamic read thresholds design, for both the uncoded and LDPC-coded flash memory channels, without any prior knowledge of the channel.

\end{abstract}


\section{Introduction}
The NAND flash memory based solid state drives have revolutionized the data storage industry. Compared with the conventional hard disk drives, they offer lower power consumption, faster write/read time, and higher reliability. However, the reliability of NAND flash memory is severely affected by various noises such as the programming noise, the random telegraph noise (RTN), and the cell-to-cell interference (CCI) \cite{cai2017error}. Most of these noises are difficult to be predicted due to the complication of memory physics. Moreover, the data retention noise caused by the charge leakage from the floating gate over time will lead to a decrease of the threshold voltages of the programmed cells \cite{cai2012error}. The corresponding change of the cell threshold voltages, referred to as the channel offset, is unknown during the readback process, and hence will severely degrade the memory sensing circuit's performance if the read thresholds still remain the same.

To mitigate the memory cell errors caused by the various noises and interference, the use of the error correction codes (ECCs) is essential. In modern solid state disks, Bose–Chaudhuri–Hocquenghem (BCH) codes with hard-decision decoding (HDD) have been employed to correct multiple bit errors \cite{lee20126, lee2014high}. To further correct errors for the multi-level cell (MLC) flash memory which is more error-prone than the single-level cell (SLC) flash memory, low-density parity-check (LDPC) codes with either HDD or soft-decision decoding (SDD) has been adopted at the cost of longer decoding latency \cite{zhao2013ldpc}.  For SDD of ECCs, the decoding performance depends heavily on the accuracy of the log-likelihood rate (LLR) of the channel coded bits, and generally more read thresholds will provide a better estimation of the LLRs of the MLC flash memory.

In this paper, we consider the MLC NAND flash memory which stores 2 bits user data per memory cell. The stored data can be differentiated by the threshold voltage of each memory cell. Specifically, a memory cell is configured to one of four threshold voltage levels with mean voltages of $V_{s_{11}}, V_{s_{10}}, V_{s_{00}}, V_{s_{01}}$, corresponding to the stored symbols of $11, 10, 00, 01$, respectively. To read the data stored in a cell, its voltage is measured and typically compared to predetermined fixed read thresholds by the memory sensing circuit. As shown by Fig. \ref{channel} (a), to differentiate four voltage levels, at least three read thresholds $a_1, a_2, a_3$ are needed. This will generate hard outputs of the MLC flash channel that can support HDD of ECCs. In order to generate the channel LLR to support SDD of ECCs such as the LDPC code, more read thresholds are required.

In the literature, several techniques have been proposed to design the read thresholds  \cite{dong2011use, wang2014enhanced,peleato2015adaptive, aslam2016read, mei2018information}. For example, a ``constant-ratio" (CR) non-uniform quantization method was proposed by observing that more errors occur in the overlapped areas of adjacent distribution functions \cite{dong2011use}. A widely accepted read thresholds design approach was reported in \cite{wang2014enhanced}, where the quantization levels are optimized by maximizing the mutual information (MMI) of the quantized channel. Based on minimizing the bit error rate (BER) and MMI, a parameter estimation method and a dynamic programming framework were proposed to design the read thresholds \cite{peleato2015adaptive}. However, all these approaches of designing the read thresholds assume that accurate threshold voltages of the memory cells can be obtained for different program/erase (P/E) cycles through proper modeling and estimation of the flash memory channel. Furthermore, they did not consider the unknown offset of the channel caused by the data retention noise. Apart from the model-based approach, read-retry and disparity-based thresholds approximation were also proposed to dynamically adjust the read thresholds such that errors can be corrected by the ECC \cite{cai2015data}. However, the read-retry scheme needs to be performed multiple times online when the ECC fails which will lead to a longer latency and more power consumption. The disparity-based method cannot ensure near-optimal performance and needs to be invoked frequently (e.g. daily) when the channel offset caused by data retention is accumulated to a large level. 

Constrained coding techniques, such as the rank modulation \cite{jiang2009rank}, balanced codes \cite{tallini1996design}, and the constant composition codes \cite{immink2017composition}, have also been proposed which can mitigate the unknown offset of the channel through sorting the channel readback signals. By leveraging on the balanced codes and the constant composition codes, the dynamic threshold schemes \cite{zhou2011error, sala2013dynamic} were proposed for both the SLC and MLC flash memories. However, a major problem with all these schemes is the high code rate loss incurred by the corresponding constrained codes.

Recently, the deep learning (DL) techniques have been developed rapidly and they have shown superior performance in many aspects of communication systems \cite{o2017introduction, gruber2017deep}. In this paper, we propose a novel DL-aided approach to design the read thresholds dynamically for the MLC flash memories. With this DL-based framework, all the unknown offset or unpredictable variations of the flash memory channel can be learned from the training data, thus avoiding the difficult task of modeling of the practical flash memories.

In particular, we first propose a novel recurrent neural network (RNN) detector to effectively detect data symbols stored in the memory cells. Compared with the prior art detection approaches, the RNN detector only requires the training data rather than an accurate model of the flash memory channel. In order to minimize the additional read latency and power consumption incurred by the RNN detector, we further propose an approach to derive the 3-level hard-decision read thresholds $a_1, a_2, a_3$ based on the output of the RNN detector. In this way, the RNN detector only needs to be activated periodically when the system is in the idle state. Once the hard-decision thresholds are obtained, the RNN detector can be terminated and the conventional threshold detector will still be adopted by using the derived hard-decision thresholds until a further adjustment of the read thresholds are needed. We name such a detector the RNN-aided (RNNA) dynamic threshold detector, which can support HDD of ECCs. Next, in order to enable SDD of ECCs, we first show how to obtain more read thresholds from our derived 3-level hard-decision read thresholds. We then propose integer-based reliability mappings based on the designed read thresholds, which can generate LLRs of the flash memory channel without any prior knowledge of the channel. Finally, to optimize the read thresholds in terms of the decoding performance, we propose to apply density evolution (DE) combined with differential evolution algorithm for LDPC coded MLC flash memory channels. We remark that our proposed RNNA dynamic read thresholds design can be easily extended to the triple-level cell (TLC) or quad-level cells (QLC) flash memories, with minor modifications of the NN parameters and the update of the integer-based reliability mappings.

The rest of the paper is organized as follows. In Section
II, we introduce the channel model for the MLC NAND flash memory adopted by this work. They are only used to generate data for training, testing, and evaluating the performance of the NN-based detector. Our proposed RNN-aided (RNNA) dynamic read thresholds design does not have any knowledge of the channel model. In Section III, we formulate the detection of the flash memory channel as a machine learning problem and propose a NN-based
detector. In Section IV, we propose novel RNNA dynamic read thresholds design, such that both the hard and soft outputs (i.e. LLRs) of the flash memory channel can be derived based on the outputs of the NN detector. Extensive
computer simulation results are illustrated in Section V. Finally, Section VI concludes the paper.

\section{Channel Model}
As extensively reported in the literature \cite{dong2011use, dong2013enabling, wang2014histogram, aslam2016read}, the threshold voltage of the flash memory cell is mainly affected by the programming noise, RTN, data retention noise, and cell-to-cell interference (CCI).

\subsection{Programming Noise}
The threshold voltage of each memory cell can be programmed by injecting certain amount of charges into the floating gate. The memory cell needs to be erased before it can be programmed. Due to process variations, the threshold voltage of erased cells is assumed to follow the  Gaussian distribution \cite{takeuchi1996double}, given by
\begin{equation}
p_{s_{11}}(v)=\frac{1}{\sqrt{2\pi}\sigma_{e}}e^{-\frac{(v-V_{s_{11}})^{2}}{2\sigma^{2}_{e}}},
\end{equation}
where $\sigma^{2}_{e}$ is the variance of the erased cell's threshold voltage. Here, $V_{s_{11}}$ denotes the mean threshold voltage of the erased state, while $V_{s_{10}}, V_{s_{00}}, V_{s_{01}}$ denote those of the programmed cells that store different symbols, respectively. To program the memory cell to these programmed voltage levels, the incremental-step-pulse programming (ISPP) \cite{lee2002effects} scheme is used, which leads to an uniform distribution of the voltage levels for programmed cells \cite{dong2011use}, given by
\begin{equation}
p_{u}(v) =   
   \begin{cases}
   \frac{1}{\bigtriangleup V_{pp}}, &\mbox{$V_p\leq v \leq V_p + \bigtriangleup V_{pp }$}\\
   0, &\mbox{otherwise},  
   \end{cases},                                                                                                                                                                                   
\end{equation}
where $V_p\in \{V_{s_{10}}, V_{s_{00}}, V_{s_{01}}\}$, and $\bigtriangleup V_{pp}$ denotes the incremental program step voltage. Moreover, the programming cells suffer from the programming noise, which is assumed to be Gaussian distributed with zero mean and variance of $\sigma^{2}_p$ \cite{dong2013enabling, wang2014histogram, dong2014using}, with $\sigma^{2}_p<\sigma^{2}_e$. Note that by using a small voltage step that realized by the programming control loop, $\sigma^{2}_p$ is significantly smaller than $\sigma^{2}_e$ \cite{compagnoni2007first}.

\subsection{Random Telegraph Noise}
In NAND flash memory, the repeated P/E cyclings lead to the  wear-out effect that damages the tunnel oxide of floating gate transistors \cite{mielke2004flash}. Such wear-out effect can be modeled by the RTN. According to \cite{compagnoni2009random}, the RTN tends to widen the voltage distributions of the memory cells and leads to exponential tails. In this paper, similar to \cite{aslam2018decision}, we model it as a Gaussian distribution for mathematical tractability, given by
\begin{equation}
p_{w}(v)=\frac{1}{\sqrt{2\pi}\sigma_{w}}e^{-\frac{v^2}{2\sigma^{2}_{w}}},
\end{equation}
where $\sigma_{w}=0.00027 N_{\text{PE}}^{0.62}$ \cite{dong2013enabling} is a function of $N_{\text{PE}}$, which denotes the number of P/E cycles.

\begin{figure}[t]
\centering
\includegraphics[scale=0.4]{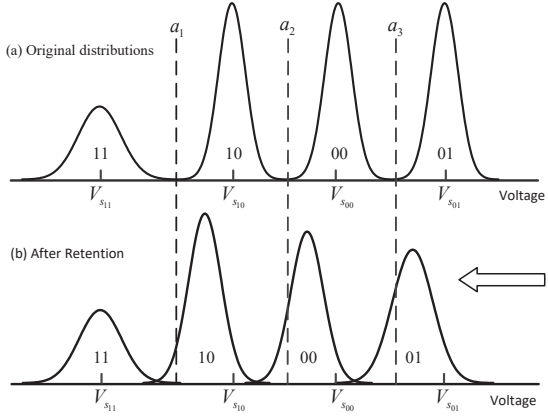}
\caption{Threshold voltage distributions of MLC flash memory cells before and after data retention noise.}
\label{channel}
\end{figure}

\subsection{Data Retention Noise}
Data retention errors are the dominant errors in flash memory. They are caused by the leakage of the charges from the floating gate over time, which leads to a decrease of the threshold voltage. Following \cite{wang2014histogram, chen2014increasing, dong2012estimating}, the retention noise is assumed to follow the Gaussian distribution, given by
\begin{equation}
p_{r_{s}}(v)=\frac{1}{\sqrt{2\pi}\sigma_{r_{s}}}e^{-\frac{(v-\mu_{r_{s}})^2}{2\sigma^{2}_{r_{s}}}},
\end{equation}
where $\mu_{r_{s}}$ and $\sigma_{r_{s}}$ are state-dependent, given by \cite{dong2013enabling}
\begin{align}
\mu_{r_{s}} &= (V_s - x_0) \cdot ( A_{t}N_{\text{PE}}^{\alpha_{i}} + B_{t}N_{\text{PE}}^{\alpha_{o}} )\cdot \ln(1+T),   \\
\sigma_{r_{s}} &= 0.3|\mu_{r_{s}}|,
\end{align}
where $s\in \{s_{11}, s_{10}, s_{00}, s_{01}\}$, $T$ is the retention time, and the other parameters $(x_0, A_{t}, B_{t}, \alpha_{i}, \alpha_{o})$ are constant. As the retention time increases, there are mainly three effects on the threshold voltage distributions of the erased and programmed states. First, the threshold voltage distributions of all states become wider. Second, the threshold voltage distributions of the programmed states shift to that of the erased state. Third, the shift of the higher-voltage states is larger than that of the lower-voltage states. 

Fig. \ref{channel} illustrates the threshold voltage distributions before and after retention noise. Here, $a_1, a_2, a_3$ are optimal hard-decision read thresholds to differentiate different states with respect to the original threshold voltage distributions. However, after data retention, it can be easily seen that these read thresholds are no longer optimal and will lead to more detection errors. Therefore, the read thresholds need to be carefully designed in order to mitigate the data retention noise induced errors.

\subsection{Overall Threshold Voltage Distributions}
The threshold voltage shift of one flash cell will influence the threshold voltage of its adjacent cells within a block, due to the parasitic capacitance-coupling effect \cite{lee2002effects}. Such interference is referred to as the CCI, which is linearly added to the threshold voltage of a victim cell. The threshold voltage shift induced by CCI can typically be compensated  by data predistortion \cite{dong2010using}, hence we assume that the CCI has already been removed in this work.

The overall threshold voltage distributions is usually approximated by the Gaussian mixture distribution \cite{cai2013threshold, lee2013estimation}, given by
\begin{equation} \label{final}
p_{s_{10} \sim s_{01}}(v)=\frac{1}{\sqrt{2\pi}\sigma_{s_{10} \sim s_{01}}}e^{-\frac{(v-\mu_{s_{10} \sim s_{01}})^2}{2\sigma^{2}_{s_{10} \sim s_{01}}}},
\end{equation}
where 
\begin{align} \nonumber
\mu_{s_{11}}&=V_{s_{11}}-\mu_{r_{s_{11}}}, \\  \nonumber
\mu_{s_{10} \sim s_{01}}&=V_{s_{10} \sim s_{01}}+\frac{\bigtriangleup V_{pp}}{2} - \mu_{r_{s_{10} \sim s_{01}}}, \\  \nonumber
\sigma_{s_{11}}^{2}&=\sigma_{e}^{2} + \sigma_{w}^{2} + \sigma_{r_{s_{11}}}^{2}, \\
\sigma^{2}_{s_{10} \sim s_{01}}&=\sigma_{p}^{2} + \sigma_{w}^{2} + \sigma_{r_{s_{10} \sim s_{01}}}^{2}.
\end{align}
In the simulations of this work, we adopt the flash memory parameters provided by \cite{dong2013enabling} and assume: $V_{s_{11}} = 1.4 $, $V_{s_{10}} = 2.6 $, $V_{s_{00}} = 3.2 $, $V_{s_{01}} = 3.93 $, $\bigtriangleup V_{pp}=0.2$, $\sigma_{e} = 0.35$, $\sigma_{p} = 0.05$, $x_0 = 1.4$, $A_t = 0.000035$, $B_t = 0.000235$, $\alpha_i=0.62$, and $\alpha_o=0.3$. With the probability density function (PDF) given in \eqref{final}, we are able to generate the readback threshold voltages for an arbitrarily long input data sequence. We remark that the channel model described above is only used to generate data for training and testing the NNs. Our subsequently proposed NN-based detection does not have any knowledge of this channel model.

\section{Learning to Detect}
In this section, we formulate the detection of the flash memory channel as a machine learning problem and propose a NN-based detector. We denote the readback threshold voltage of the $k$-th memory cell by $v_k$. The inputs to the NN are $\bm{v}=\left\lbrace v_1, v_2, \cdots, v_L \right\rbrace $, where $L$ is the number of neurons in the input layer of the NN. The outputs of the NN are the soft estimates $\bm{\tilde{x}}=\left\lbrace \tilde{x}_1, \tilde{x}_2, \cdots, \tilde{x}_L \right\rbrace$ of the labels $\bm{x}$. For MLC flash memory, there are four different states $V_{s_{11}}, V_{s_{10}}, V_{s_{00}}, V_{s_{01}} $, which can be represented by labels $\{ 0, 1 ,2 ,3 \}$, respectively. Hence, for the $k$-th cell, $x_{k} \in \{ 0, 1 ,2 ,3 \}$, the hard estimate $\hat{x}_{k} \in \{ 0, 1 ,2 ,3 \}$ can be obtained by taking the nearest integer of $\tilde{x}_k$. Thereafter, the corresponding recovered bits $c_{2k-1}$ and $c_{2k}$ can be obtained by using the mapping $\{ 0, 1 ,2 ,3 \} \rightarrow \{ 11, 10 , 00 ,01 \}$.

The outputs $\bm{\tilde{x}}$ of the NN can be considered as a function of the NN's inputs $\bm{v}$ and network's parameters $\bm{\theta}$, given by $\bm{\tilde{x}}=f(\bm{v}, \bm{\theta})$. Then, the NN is trained to find the best $\bm{\theta^{*}}$ that leads to a well-performed detector. To accomplish this task, a loss function $\mathcal{L}$ is defined over the set of training data, such that 
\begin{equation}
\bm{\theta^*}=\text{arg}\min_{\bm{\theta}}\mathcal{L}(\bm{x},\bm{\tilde{x}}),
\end{equation}
where $\mathcal{L}(\bm{x},\bm{\tilde{x}})$ measures the loss between $\tilde{\bm{x}}$ and $\bm{x}$. By using the gradient descent algorithm or its variants, combined with the back propagation method, $\bm{\theta^*}$ can be obtained by minimizing $\mathcal{L}(\bm{x},\bm{\tilde{x}})$ defined above over the training data set.

Note that as a supervised learning approach, the labels $\bm{x}$ are known during the training. After the training process, the network parameters $\bm{\theta^*}$ is obtained and another data set named the validation set will be used to finetune the model hyperparameters. After training and validation, the NN detector will be employed to detect the unknown flash memory channel outputs using the pre-determined NN.

\subsection{Neural Network Architecture}
\begin{figure}[t]
\centering
\includegraphics[height=2.2in,width=3.1in]{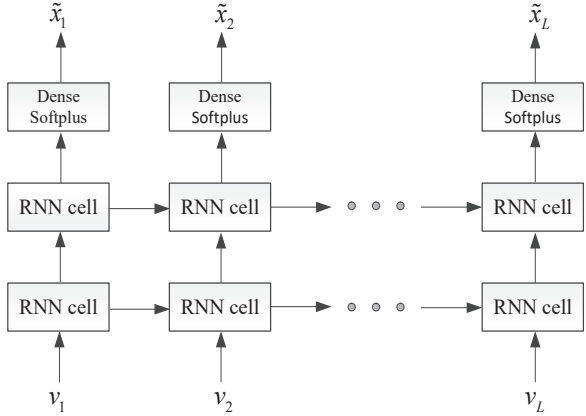}
\caption{Proposed RNN architecture for the NN-based  detection.}
\label{RNN}
\end{figure}

In this paper, we propose to use a stacked RNN architecture to perform the flash memory channel detection. The RNN is a class of NNs with feedback connections. It is very suitable to time series tasks, since it can use memories to process sequences of inputs. There are different types of RNN cells, namely, the vanilla RNN, gated recurrent unit (GRU) and long short-term memory (LSTM). The number of network parameters in vanilla RNN is significantly less than that of the GRU and LSTM. However, it suffers from the vanishing/exploding gradient problem, where the gradients may vanish to zero or explode during the training process \cite{goodfellow2016deep}. Hence, we employ the GRU as the RNN cell, since it has less parameters than the LSTM. Fig. \ref{RNN} shows the proposed RNN architecture with two GRU layers and one fully-connected output layer. For each layer, an activation function is applied to introduce the non-linearity to the NN. For the two GRU hidden layers, the rectified linear unit (ReLU) activation function is adopted, given by $\sigma_{\text{relu}}(t)=\max\left\lbrace 0,t \right\rbrace$ with $\sigma_{\text{relu}}(t)\in [0, \infty)$ . For the output layer, to obtain soft estimates of $\bm{x}$, we use the softplus activation function, which is the smooth approximation of the ReLU function, given by $\sigma_{\text{softplus}}(t)=\ln(1+\exp(x))$ with $\sigma_{\text{softplus}}(t)\in [0, \infty)$.

\subsection{Training Approach}
To train the RNN, the readback threshold voltages $\bm{v}$ can be obtained by sensing the current memory cell. In our work, we generate the readback threshold voltages by simulating the flash channel model described in Section II. As a supervised learning approach, the label $x_k$ of $v_k$ is also required for the training the RNN. To solve this problem, we use all codewords that are decoded by the LDPC decoder as labels to train the RNN until a desired number of labels are collected. This method may lead to a performance degradation due to the existence of codewords that are incorrectly decoded by the LDPC code. Note that it is possible to collect more accurate labels by checking the syndrome of the decoded codeword, but at the cost of additional latency. The performance comparison of the RNN detector with correct and error corrupted labels will be presented in Section V. 

For the proposed RNN, to minimize the difference between NN output $\tilde{\bm{x}}$ and labels $\bm{x}$, we define the loss function as the mean square error (MSE) between $\tilde{\bm{x}}$ and $\bm{x}$, given by 
\begin{equation}
 \mathcal{L}(\bm{x},\bm{\tilde{x}})=\frac{1}{L}\sum_{k=1}^{L} (x_k-\tilde{x}_k)^2.
\end{equation} 
In our experiments, the number of neurons $L$ in the input layer is set to be 50 and this number can be reduced to further simplify the network. After many trials, it is found that $ 3\times 10^{6}$ training symbols are sufficient for the RNN to achieve its best performance. For example, for an LDPC code with codeword length $8000$ bits, $750$ recovered codewords are enough for training the RNN.

\begin{figure}[t] 
\centering
\includegraphics[height=2.1in,width=3.2in]{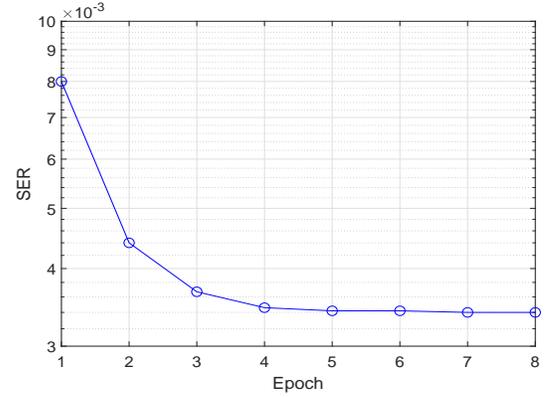}
\caption{The training SER of the RNN detector for each epoch at P/E cycles $N_{\text{PE}} = 10^{4}$ and retention time $T=10^{2}$ hours.}
\label{RNN_train}
\end{figure}

The details of the RNN setting are summarized in Table I. To illustrate the learning process, Fig. \ref{RNN_train} shows the symbol error rate (SER) of the RNN detector with correct labels for each epoch during training. It is observed that the training SER decreases sharply at the second epoch and converges from the fourth epoch. It shows that the RNN can learn the channel variations within only several epochs.

We remark that the proposed NN detection framework is data-driven, which does not need the prior knowledge of the channel. After training and validation, the proposed RNN detector can successfully detect the readback signal $\bm{v}$. Furthermore, the above proposed NN detection framework can be easily extended to the TLC or QLC flash memories, with minor modifications to the NN parameters. For example, more training data and hidden layers may be required to successfully detect the readback signal.

\renewcommand\arraystretch{1.4}
\begin{table}[htbp]
\centering
\caption{Network Settings for the Proposed RNN Detector.}
\begin{tabular}{|c|c|c|}
\hline
Training Symbols & $3\times 10^{6}$           \\ \hline
Mini-batch Size  & $100$                      \\ \hline
Loss Function    & MSE                        \\ \hline
Initializer      & Xavier uniform initializer \\ \hline
Optimizer        & Adam optimizer             \\ \hline
\end{tabular}
\end{table}

\section{Neural Network-Aided Dynamic Read Thresholds Design}
As will be illustrated in Section V, the proposed NN detector can achieve near-optimal performance without any knowledge of the channel. However, it needs to be activated for each data block of length $L$ for data detection. Although the NN can be efficiently implemented in parallel with powerful hardware such as the graphical processing units (GPU) or the application-specific integrated circuit (ASIC), it will still lead to substantially additional read latency and power consumption. To avoiding using NN to detect every block of data, in this section, we propose novel NNA dynamic read thresholds design, such that both the hard and soft outputs (i.e. LLRs) of the flash memory channel can be derived based on the outputs of the NN detector.


\subsection{Read Thresholds Design for Generating Hard Channel Outputs}
For the MLC flash memory channel, in order to generate hard-outs of the channel to support HDD of ECCs, 3 read threshold levels $\{a_1, a_2, a_3\}$ need to be determined. For a given $\bm{v}$ and with assumed $\{a_1, a_2, a_3\}$, we can obtain the hard estimate $\bm{\bar{x}}$. Meanwhile, based on $\bm{v}$, the RNN outputs $\bm{\tilde{x}}$ and hence $\bm{\hat{x}}$. Therefore, the adjusted thresholds $\{a^{*}_1, a^{*}_2, a^{*}_3\}$ can be obtained by searching for the thresholds that can minimize the Hamming distance between $\bm{\bar{x}}$ and $\bm{\hat{x}}$, denoted by $d(\bm{\bar{x}}, \bm{\hat{x}})$. We assume that $M$ RNN output sequences with length $L$ are involved in the searching and denote them by $\hat{\bm{X}}=\{\hat{\bm{x}}_1, \hat{\bm{x}}_2, \ldots, \hat{\bm{x}}_M  \}$. We further have $\bar{\bm{X}}=\{\bar{\bm{x}}_1, \bar{\bm{x}}_2, \ldots, \bar{\bm{x}}_M  \}$ and ${\bm{V}}=\{{\bm{v}}_1, {\bm{v}}_2, \ldots, {\bm{v}}_M  \}$. Hence, we can obtain
\begin{equation} \label{a123}
\{a^{*}_1, a^{*}_2, a^{*}_3\} = \mathop{\arg\min}_{\{ a_1, a_2, a_3\}} d(\bar{\bm{X}}, \hat{\bm{X}}).
\end{equation}

To get $\{a^{*}_1, a^{*}_2, a^{*}_3\}$, we first uniformly quantize the search space into $m$ intervals (e.g. $ m=1000$), with boundaries $b_0, b_1, \ldots, b_m$ where $b_0=-\infty< b_1=V_{s_{11}} < \cdots < b_{m-1}=V_{s_{01}} < b_m=\infty$. Note that larger $m$ will result in higher precision, but at the cost of higher computational complexity. Hence, we have $a_{i}\in\{ b_0, b_1, \ldots, b_m\}$, $i=0,1,\ldots,n$ with $a_0=-\infty$ and $a_n=\infty$. In our case, we have $n=4$, since there are three read thresholds. To solve the problem in \eqref{a123} efficiently, a precomputation step is performed first. In this step, the elements in $\bm{V}$ are first sorted in ascending order with complexity $\mathcal{O}(ML\log(ML))$. Based on the corresponding $\hat{\bm{X}}$ obtained at the output of the RNN, we can get the number of each symbol $i$ that falls into $[b_{j}, b_{k})$, denoted by $s(i, b_j, b_k)$, with $i=0,1,\ldots,n-1$, $0\leq j<k\leq m$. Hence, \eqref{a123} is equivalent to
\begin{equation} \label{astar}
\{a^{*}_1, \ldots, a^{*}_{n-1}\} = \mathop{\arg\min}_{\{a^{*}_1, \ldots, a^{*}_{n-1}\}} \left(  ML - \sum_{i=0}^{n-1} s(i,a_{i}, a_{i+1}) \right)  .
\end{equation}
Note that \eqref{astar} can be further simplified to
\begin{equation} \label{astar_sim}
\{a^{*}_1, \ldots, a^{*}_{n-1}\} = \mathop{\arg\max}_{\{a^{*}_1, \ldots, a^{*}_{n-1}\}} \sum_{i=0}^{n-1} s(i,a_{i}, a_{i+1})  .
\end{equation}
Then, an exhaustive search method can be used to find $a^{*}_1, \ldots, a^{*}_{n-1}$ from $b_1, b_2, \ldots, b_{m-1}$ with computational complexity $\mathcal{O}(m^{n-1})$. To further reduce the complexity, we propose to apply dynamic programming (DP) \cite{Cormen} to solve \eqref{astar_sim} through a method similar to \cite{he2019dynamic}. That is, let $P(m',n')$ ($1<n'\leq m'\leq m$,) denote the problem of finding $a_1, a_2, \ldots, a_{n'-1}$ from $b_1, b_2, \ldots, b_{m'-1}$ such that $C(m',n')$ is maximized, where $C(m',n') = \sum_{i=0}^{n'-1} s(i,a_{i}, a_{i+1}) $ is the objective function in \eqref{astar_sim}. We denote $C^{*}(m,n)$ as the objective function of the optimal solution of $P(m, n)$. Hence, $C^{*}(m,n)$ is given by
\begin{align} \label{recursion} \nonumber
&C^{*}(m, n) \\ \nonumber
& = \mathop{\max}_{n-1\leq\lambda_{n-1}<m} \left[  C^{*}(\lambda_{n-1}, n-1) + s(n-1,b_{\lambda_{n-1}}, b_{m}) \right] , \\ 
& = C^{*}(\lambda^{*}_{n-1}, n-1) + s(n-1, b_{\lambda^{*}_{n-1}}, b_{m}),
\end{align}
where $\left\lbrace \lambda^{*}_{1},\ldots, \lambda^{*}_{n-1}\right\rbrace $ collects the indices of $b_{1} , \ldots, b_{m-1}$ and $\{b_{\lambda^{*}_{1}} , \ldots, b_{\lambda^{*}_{n-1}}\}$ is the optimal solution of $P(m, n)$. From \eqref{recursion}, the optimal solution of $P(m, n)$ can be obtained by solving its subproblems $P(\lambda_{n-1}, n-1)$, where $n-1\leq\lambda_{n-1}<m$. Similarly, $P(\lambda_{n-1}, n-1)$ can also be solved by its subproblems $P(\lambda_{n-2}, n-2)$, where $n-2\leq \lambda_{n-2}<\lambda_{n-1}$. In this way, the optimal solution of $P(m, n)$ can be calculated in a recursive manner, such that DP can be employed \cite{Cormen}. According to \eqref{recursion}, the complexity of DP is given by $\mathcal{O}(m^{2}n)$. Since $n\ll m$, the complexity of DP is much lower than the exhaustive search method. Note that when extending this work to TLC or QLC flash memories, the exhaustive search method is prohibitive since its complexity is $\mathcal{O}(m^{n-1})$, with $n=2^3$ and $n=2^4$ for TLC and QLC flash memories, respectively. On the other hand, DP is always feasible to solve this problem, since its complexity is always linearly proportional to $m^2$.

\begin{figure*}
\begin{align} \label{SEP} \nonumber
P_{s}&= \sum_{i} P(v = V_{s_{i}})P(e|v = V_{s_{i}})                \\        \nonumber
&= \frac{1}{4}\big(  P(v>a_1 | v=V_{s_{11}}) +P(v<a_1 \cup v>a_2  | v=V_{s_{10}}) + P(v<a_2 \cup v>a_3  | v=V_{s_{00}})+               P(v<a_3 | v=V_{s_{01}}) \big) \\
&= \frac{1}{4} \left\lbrace   3 +Q\left( \frac{a_1-\mu_{s_{11}}}{\sigma_{s_{11}}} \right)  - Q\left( \frac{a_1-\mu_{s_{10}}}{\sigma_{s_{10}}} \right) + Q\left( \frac{a_2-\mu_{s_{10}}}{\sigma_{s_{10}}} \right)  - Q\left( \frac{a_2-\mu_{s_{00}}}{\sigma_{s_{00}}} \right) + Q\left( \frac{a_3-\mu_{s_{00}}}{\sigma_{s_{00}}} \right)  - Q\left( \frac{a_3-\mu_{s_{01}}}{\sigma_{s_{01}}} \right) \right\rbrace .
\end{align}
\hrule
\end{figure*}

We remark that the RNN detection and the subsequent search of the read thresholds are only activated periodically when the system is in the idle state, and will be terminated once $\{a^{*}_1, a^{*}_2, a^{*}_3\}$ are obtained. Thereafter, the threshold detector with the obtained read thresholds will still be adopted until a further adjustment of the read thresholds is necessary. For convenience, we name such a detector the RNNA dynamic threshold detector. Hence, compared with the RNN detection that is carried out for each data block, our proposed RNNA dynamic threshold detector leads to a significant reduction of the read latency and power consumption.

To evaluate the performance of the proposed RNN detector and the RNNA dynamic threshold detector, we derive the optimum symbol error probability (SEP) of the MLC flash memory channel with the full channel knowledge, and use it as the performance benchmark. Assuming four voltage states $\{V_{s_{11}}, V_{s_{10}}, V_{s_{00}}, V_{s_{01}} \}$ are equiprobably stored in the memory cells. Given read thresholds $\{a_1, a_2, a_3\}$ and based on the channel model described in Section II, the SEP is given by \eqref{SEP}, where $e$ is a symbol error and $i\in\{11, 10, 00, 01   \}$. When Gray mapping is used, the adjacent states only differ with 1 bit. Hence, the corresponding bit error probability (BEP) can be approximated as $P_b\approx 0.5P_s$.

\par The SEP in \eqref{SEP} is a function of $\{a_1, a_2, a_3\}$, which is a continuous function and is locally quasi-convex within the range of our interest. Hence, we can find the optimum read thresholds, and hence the minimum SEP by using the Newton-Raphson method or other quasi-convex optimization techniques \cite{boyd2004convex}. The derived minimum SEP can serve as the lower bound to evaluate the proposed detectors, whose performance will be illustrated in Section V.


\subsection{Read Thresholds Design for Generating Soft Channel Outputs}
The 3-level hard-decision read thresholds derived above can be used to differentiate four types of symbols stored in the MLC flash memory cell. In order to performance SDD, more read thresholds are needed. In this subsection, we first show how to design more read thresholds to support SDD by using the 3-level hard-decision read thresholds designed earlier. We use the 6-level read thresholds as an example to generate the soft channel outputs, although the proposed thresholds design method can be generalized to more number of read thresholds. We then propose integer-based reliability mappings based on the designed read thresholds, which can generate LLRs of the flash memory channel without any prior knowledge of the channel. Moreover, to optimize the read thresholds in terms of the decoding performance, we propose to apply DE combined with differential evolution algorithm for LDPC coded MLC flash memory channels.


\begin{figure}[t]
\centering
\includegraphics[height=1.2in,width=3.4in]{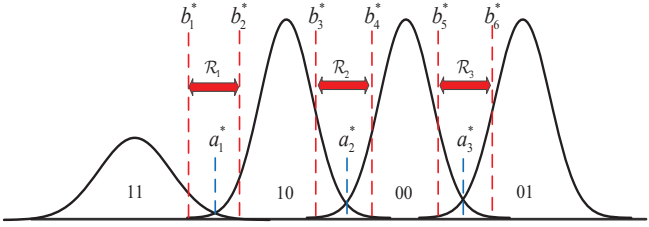}
\caption{Defining the non-uniform read thresholds $b_{1}^{*}, b_{2}^{*}, \ldots, b_{6}^{*}$ based on $a_{1}^{*}, a_{2}^{*}, a_{3}^{*}$.}
\label{quan}
\end{figure}

\begin{table*}[t]
\centering
\caption{The Proposed Mappings Between the quantization level that the channel readback signal belongs to and the Integer-based Reliability}
\begin{tabular}{|c|c|c|c|c|c|c|c|}
\hline
                 & $[b_0^{*}, b_1^{*})$ & $[b_1^{*}, b_2^{*})$ & $[b_2^{*}, b_3^{*})$ & $[b_3^{*}, b_4^{*})$ & $[b_4^{*}, b_5^{*})$ & $[b_5^{*}, b_6^{*})$ & $[b_6^{*}, b_7^{*})$ \\ \hline
$\hat{L}_{\text{MSB}}$ & $-3$           & $-2$           & $-1$           & $0$            & $1$            & $2$            & $3$            \\ \hline
$\hat{L}_{\text{LSB}}$ & $-1$           & $0$            & $1$            & $2$            & $1$            & $0$            & $-1$           \\ \hline
\end{tabular}
\end{table*}


\subsubsection{Soft-Decision Read Thresholds Design}

As illustrated by Fig. \ref{quan}, in flash memories, the  dominant overlapping regions of two adjacent distribution functions are around the hard-decision read thresholds, where more errors will occur. Hence, it is natural to sense this region with a higher precision and the remaining region with a lower precision. Therefore, we propose to adopt non-uniform read thresholds $\{b_{1}^{*}, b_{2}^{*}, \ldots, b_{6}^{*}\}$ based on the hard-decision read thresholds $\{a_{1}^{*}, a_{2}^{*}, a_{3}^{*}\}$ we derived earlier. In particular, to find $\{b_{1}^{*}, b_{2}^{*}, \ldots, b_{6}^{*}\}$, we first define three regions $\mathcal{R}_{1}$, $\mathcal{R}_{2}$, $\mathcal{R}_{3}$, whose centers are $a_{1}^{*}, a_{2}^{*}, a_{3}^{*}$, respectively. Assume the widths of $\mathcal{R}_{1}$, $\mathcal{R}_{2}$, $\mathcal{R}_{3}$ are $\mathcal{W}_1$, $\mathcal{W}_2$, $\mathcal{W}_3$, respectively. Then, we propose to obtain the boundaries $\{b_{1}^{*}, b_{2}^{*}, \ldots, b_{6}^{*}\}$ by
\begin{equation} \label{bi}
   \begin{cases}
b_{2i-1}^{*} =  a_{i}^{*} - \mathcal{W}_{i}/2,   \\
b_{2i}^{*} = a_{i}^{*} + \mathcal{W}_{i}/2,
   \end{cases},
\end{equation}
where $a_{i}^{*}$ is the center of each interval $[b_{2i-1}^{*}, b_{2i}^{*}]$ with $i=1,2,3$. In this way, by using \eqref{bi}, the optimization of $\{b_{1}^{*}, b_{2}^{*}, \ldots, b_{6}^{*}\}$ is converted to the problem of optimization of $\{\mathcal{W}_1$, $\mathcal{W}_2$, $\mathcal{W}_3\}$, which is a much easier task. The brute force way to optimize $\mathcal{W}_1$, $\mathcal{W}_2$ and $\mathcal{W}_3$ is by computer exhaustive search through Monte-Carlo simulations so as to minimize the error rate performance of the system. However, for specific ECCs, such as the LDPC code, we can use theoretical analysis to replace the error rate simulations for optimizing the values of $\mathcal{W}_1$, $\mathcal{W}_2$ and $\mathcal{W}_3$. The corresponding details are presented in Section IV.B.3).

\subsubsection{Integer-based Reliability Mappings}
With the designed read thresholds, the next step is to derive LLRs for SDD. Frist, as a reference, we show how to obtain the LLR with the designed read thresholds $\{b_{1}^{*}, b_{2}^{*}, \ldots, b_{6}^{*}\}$, by using the full knowledge of the channel. Each memory cell has four possible voltage states and stores two bits, we refer the left bit as the most significant bit (MSB) and the right bit as the least significant bit (LSB). Define $T_j = [b_j, b_{j+1})$ as the $j$-th quantization interval, $j=0,1,\ldots,6$, with $b_{0}=-\infty$ and $b_{7}=\infty$. Then, for a given threshold voltage $v\in T_j$ and assume the channel PDF of \eqref{final} is known, the LLR of the MSB and the LSB are given by
\begin{align} \nonumber \label{MSB}
L_{\text{MSB}}&=\ln \frac{\text{Pr}(v\in T_j | \text{MSB} = 0)}{\text{Pr}(v\in T_j | \text{MSB} = 1)} \\
&= \ln\frac{\int_{T_j}\{p_{s_{00}}(v) + p_{s_{01}}(v) \}dv }{\int_{T_j}\{p_{s_{10}}(v) + p_{s_{11}}(v) \}dv}
\end{align}
and
\begin{align} \nonumber \label{LSB}
L_{\text{LSB}}&=\ln \frac{\text{Pr}(v\in T_j | \text{LSB} = 0)}{\text{Pr}(v\in T_j | \text{LSB} = 1)} \\
&= \ln\frac{\int_{T_j}\{p_{s_{00}}(v) + p_{s_{10}}(v) \}dv }{\int_{T_j}\{p_{s_{01}}(v) + p_{s_{11}}(v) \}dv},
\end{align}
where
\begin{align} \nonumber
\int_{T_j}p_{s_{10} \sim s_{01}}(v)dv &= Q\left( \frac{b_{j}-\mu_{s_{10} \sim s_{01}}}{\sigma_{s_{10} \sim s_{01}}}  \right)\\
&-Q\left( \frac{b_{j+1}-\mu_{s_{10} \sim s_{01}}}{\sigma_{s_{10} \sim s_{01}}}  \right).
\end{align}

However, in practical flash memories, due to the various noises/interference and especially the data retention noise, the accurate PDF of each voltage state is not available to the channel detector. Hence, accurate calculation of the LLRs given by \eqref{MSB} and \eqref{LSB} is not possible. However, it has been shown that the reliability of a received symbol can be measured by its magnitude and this measure can be quantized to integers for the additive white Gaussian noise (AWGN) channel \cite{chen2011comparisons} and the single-level-cell NVM channel \cite{cai2013channel}. In this paper, we propose integer-based reliability mappings for SDD of ECCs over the MLC flash memory channel.

In particular, as shown by Fig. \ref{quan}, in the dominant overlapped regions $\mathcal{R}_{2}$, we are the least confident about whether the MSB is a `0' or a `1'. Hence, if the readback signal $v\in \mathcal{R}_{2}$, set $\hat{L}_{\text{MSB}}=0$. Similarly, if $v\in \mathcal{R}_{1}$ or $v\in\mathcal{R}_{3}$, set $\hat{L}_{\text{LSB}}=0$. If $v$ is farther away from these dominant overlapped regions, we are more confident about whether the MSB is a `0' or a `1'. Similar trend holds for the LSB. Based on these observations, it is natural to given a mapping between the quantization level that the channel readback signal belongs to, and the integer-based reliability $\hat{L}_{\text{MSB}}$ ($\hat{L}_{\text{LSB}}$) as shown by Table II. Note that the mappings we propose in Table II may be not optimal and better mappings could be found to further enhance the performance.

\subsubsection{LDPC Code-Specific Optimization of the Read Thresholds }

With the above proposed mappings, the SDD of ECCs can then be performed. In this work, we consider the LDPC codes, which have already been widely applied to MLC flash memory channels \cite{cai2017error}. Hence the above proposed integer-based reliability measure can be fed into the decoder of the LDPC codes. In this work, the normalized min-sum (NMS) decoding algorithm is employed since it can closely approach the performance of the sum-product algorithm (SPA) with a much lower computational complexity. 

In this subsection, to optimize the LDPC-coded performance over the MLC flash memory channel, $\mathcal{W}_1$, $\mathcal{W}_2$ and $\mathcal{W}_3$ are jointly optimized for the LDPC codes. To evaluate the decoding performance, the DE analysis can be employed to derive the decoding threshold of LDPC ensembles. The ensemble of LDPC codes can be characterized by the degree distributions. A regular LDPC ensemble is defined by $(d_v, d_c)$, where $d_v$  and $d_c$ are the number of edges connected to each variable node and check node, respectively. An irregular LDPC ensemble has non-uniform variable node and check node degrees, hence it can be defined by edge degree distributions $\lambda(x)$ and $\rho(x)$, given by
\begin{equation}
\lambda(x)=\sum_{j\geq 2}\lambda_{j}x^{j-1} ,\quad \rho(x)=\sum_{i\geq 2}\rho_{i}x^{i-1},
\end{equation} 
where $\lambda_{j}$ and $\rho_{i}$ are the fraction of edges that are connected to variable and check nodes with degree $j$ and $i$, respectively.

Since the MLC flash memory channel is asymmetric, to enable DE, the channel symmetrizing method proposed in \cite{richardson2008modern} is employed. First, for a given set of widths $\mathcal{W}_1$, $\mathcal{W}_2$, $\mathcal{W}_3$ of $\mathcal{R}_{1}$, $\mathcal{R}_{2}$, $\mathcal{R}_{3}$, the integer-based reliability $\hat{L}_{\text{MSB}}$ ($\hat{L}_{\text{LSB}}$) of the readback signal $v$ can be obtained based on the mappings given by Table II. We then define $f_{0}(\hat{L})$ and $f_{1}(\hat{L})$ as the PDFs of $\hat{L}$ corresponding to the originally stored bit of $c = 0$ and $c = 1$, respectively. Note that $f_{0}(\hat{L})$ and $f_{1}(\hat{L})$ can be obtained using a histogram approach.

According to the channel symmetrizing method \cite{richardson2008modern}, we can flip all the signs of
LLRs with $x = 0$. Due to the symmetry of the processing rules,  the signs of messages that enter or exit the variable nodes are also flipped. Hence, the DE for a particular codeword is equivalent to that for the all-one codeword. The $L$-density after channel symmetrizing is given by
\begin{equation}
f_{s}(\hat{L})=\frac{1}{2}\left( f_{0}(-\hat{L})+ f_{1}(\hat{L}) \right).
\end{equation}
Then, the $L$-density $f^{s}(\hat{L})$ can be used to initialize the DE algorithm. The DE of the NMS algorithm is an iterative algorithm which consists of the evolution of the $L$-densities for the check node and variable node updates. The original DE of the NMS algorithm is presented in \cite{chen2005reduced}. In this work, we employ the discrete DE (DDE) \cite{chung2001design} to reduce the complexity of DE by quantizing all input and output messages during decoding. Let the $u^{(l)}$ be the message from a degree-$d_c$ check node to a variable node at the $l$-th iteration, and $v^{(l)}$ be the output message of a degree-$d_v$ variable node at the $l$-th iteration. Under the NMS algorithm with normalization factor $\alpha$, the check node update and variable node update are given by
\begin{align}  \nonumber
u^{(l)} &=  \left( \prod_{j=1}^{d_c-1} \text{sign} \left( v^{(l-1)}_j \right)\right)  \cdot \alpha \cdot \min_{j\in\{1,2,\ldots, d_c-1\}} |v^{(l-1)}_j|,         \\ 
v^{(l)} &= v^{(0)} + \sum_{i=1}^{d_v-1}u^{(l)}_{i},
\end{align}
respectively, where $\alpha$ is the normalization factor, $v^{(l)}_j$, $j=1,2,\ldots,d_c-1$ are the incoming messages from neighbors of a degree-$d_c$ check node, and $u^{(l)}_i$ , $i=1,2,\ldots,d_v-1$ are the incoming messages from neighbors of a degree-$d_v$ variable node at the $l$-th iteration. The initial message of the algorithm is $v^{0}=\hat{L}$. According to the quantized NMS algorithm, $u^{(l)}$ and $v^{(l)}$ are quantized to $\bar{u}^{(l)}$ and $\bar{v}^{(l)}$. We denote the probability mass function (PMF) of $\bar{u}^{(l)}$ and $\bar{v}^{(l)}$ by $P_{\bar{u}}^{(l)}$ and $P_{\bar{v}}^{(l)}$, and they can be calculated using the DDE approach given in \cite{chung2001design}. According to the DDE algorithm, the fraction of incorrect messages
for the $l$-th iteration is given by
\begin{equation} \label{pe}
P_e^{(l)}=\sum_{k<0} P_{\bar{v}}^{(l)}[k].
\end{equation}

\begin{figure}[t] 
\centering
\includegraphics[height=2.4in,width=3.6in]{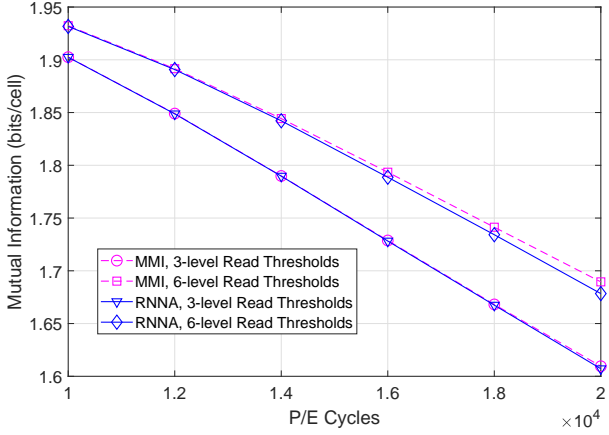}
\caption{The MI with the MMI quantizer and RNNA quantizer over different P/E cycles at $T=10^{4}$ hours.}
\label{MI}
\end{figure}

As described earlier, the initial integer-reliability $\hat{L}$, and hence the $L$-density is determined by $\mathcal{W}_1$, $\mathcal{W}_2$, $\mathcal{W}_3$. Therefore, the choice of $\mathcal{W}_1$, $\mathcal{W}_2$ and $\mathcal{W}_3$ will directly affect $P_e^{(l)}$ given by \eqref{pe}. Differential evolution is an global optimization algorithm that does not rely on any assumption of the problem \cite{storn1997differential}. In this work, we apply the differential evolution algorithm to find the a set of $\mathcal{W}_1$, $\mathcal{W}_2$ and $\mathcal{W}_3$ so as to optimize the decoding performance. The cost function is $P_e^{(l)}$ and hence the optimized $\{\mathcal{W}_1^{*}, \mathcal{W}_2^{*}, \mathcal{W}_3^{*}\}$ are given by
\begin{equation}
\{\mathcal{W}_1^{*}, \mathcal{W}_2^{*}, \mathcal{W}_3^{*}\} = \mathop{\arg\min}_{\{\mathcal{W}_1, \mathcal{W}_2, \mathcal{W}_3\}} P_e^{(l)},
\end{equation}
with given channel parameters $N_{\text{PE}}$, $T$ and the number of iterations (e.g. $l=10$). The details of differential evolution optimization process are presented in \cite{storn1997differential}.

After deriving the optimized $\mathcal{W}_1^{*}, \mathcal{W}_2^{*}, \mathcal{W}_3^{*}$, the corresponding read thresholds $\{b_{1}^{*}, b_{2}^{*}, \ldots, b_{6}^{*}\}$ can be obtained according to \eqref{bi}. We remark that for extending this work to TLC or QLC flash memories, the above described RNNA dynamic read thresholds design method is still valid. The only difference lies in the design of the mappings between the integer based reliability and the quantization level that the channel readback signal belongs to, since more read thresholds are required for TLC and QLC flash memories.

Fig. \ref{MI} compares the MI of the flash memory channel with the MMI quantizer \cite{wang2014enhanced} and the RNNA quantizer proposed in this subsection. It is observed that the MI with the proposed RNNA quantizer is very closed to that with the MMI quantizer for both the 3-level and 6-level read thresholds. Note that that for the case with the MMI quantizer, it is assumed that threshold voltage distributions of the flash memory channel is known which is unrealistic, while for the case with the RNNA quantizer, there is no prior knowledge of the channel at all, which is consistent with the practical flash memories.

\begin{figure}[t] 
\centering
\includegraphics[height=2.4in,width=3.6in]{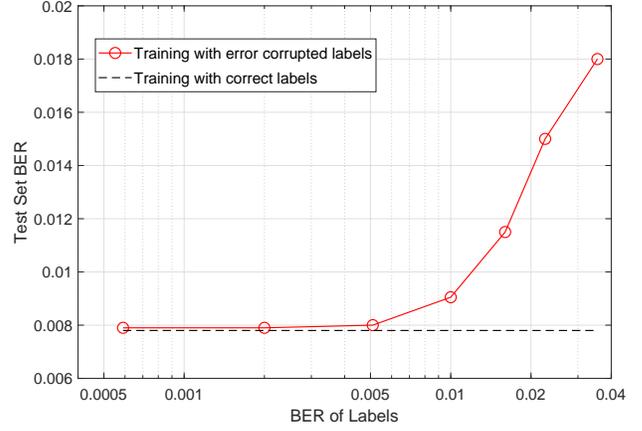}
\caption{The test set BER using correct and error corrupted labels respectively at $N_{\text{PE}}=1.1\times 10^{4}$ and $T=10^{4}$ hours.}
\label{error_label}
\end{figure}

\begin{figure}[t] 
\centering
\includegraphics[height=2.4in,width=3.6in]{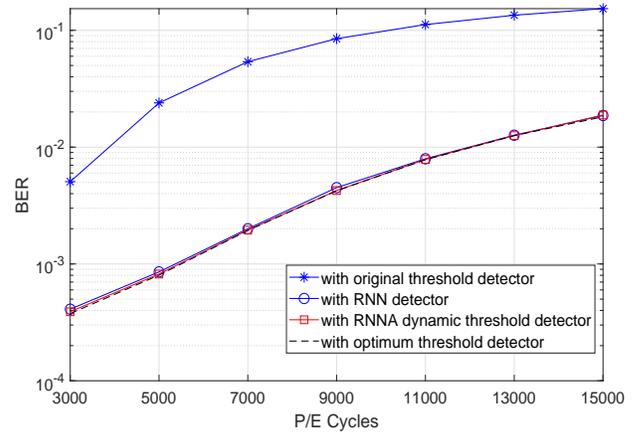}
\caption{BER of the RNN detector and RNNA dynamic threshold detector at $T=10^{4}$ hours.}
\label{ser_d}
\end{figure}

\begin{figure}[t] 
\centering
\includegraphics[height=2.4in,width=3.6in]{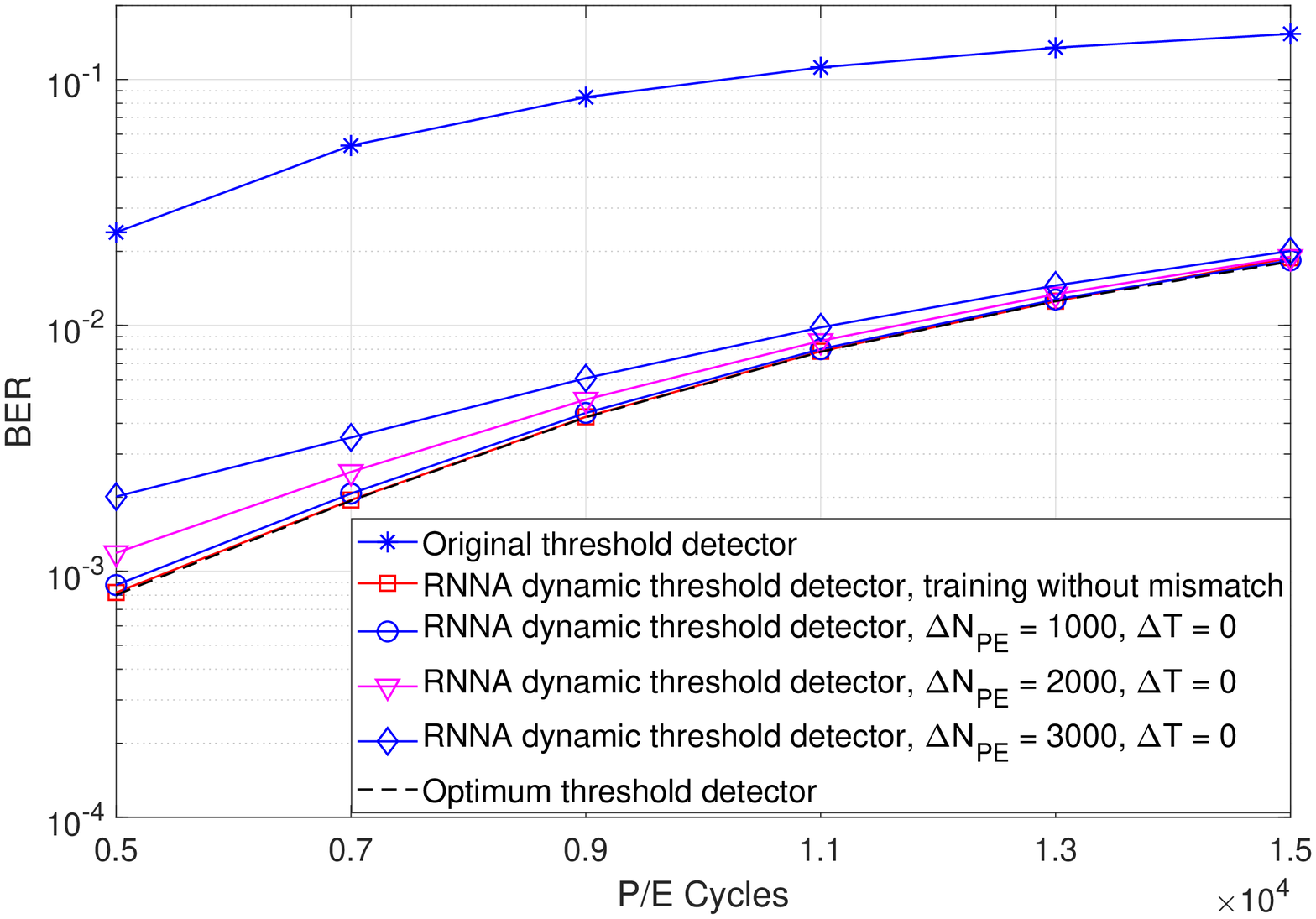}
\caption{BER comparison of different detectors. The RNNA dynamic threshold detector is trained with P/E cycles mismatch $(\bigtriangleup N_{\text{PE}}=1000, 2000, 3000)$ and at $T^{\text{test}}=10^{4}$ hours.}
\label{ser_mismatch_pe}
\end{figure}


\begin{figure}[t] 
\centering
\includegraphics[height=2.4in,width=3.6in]{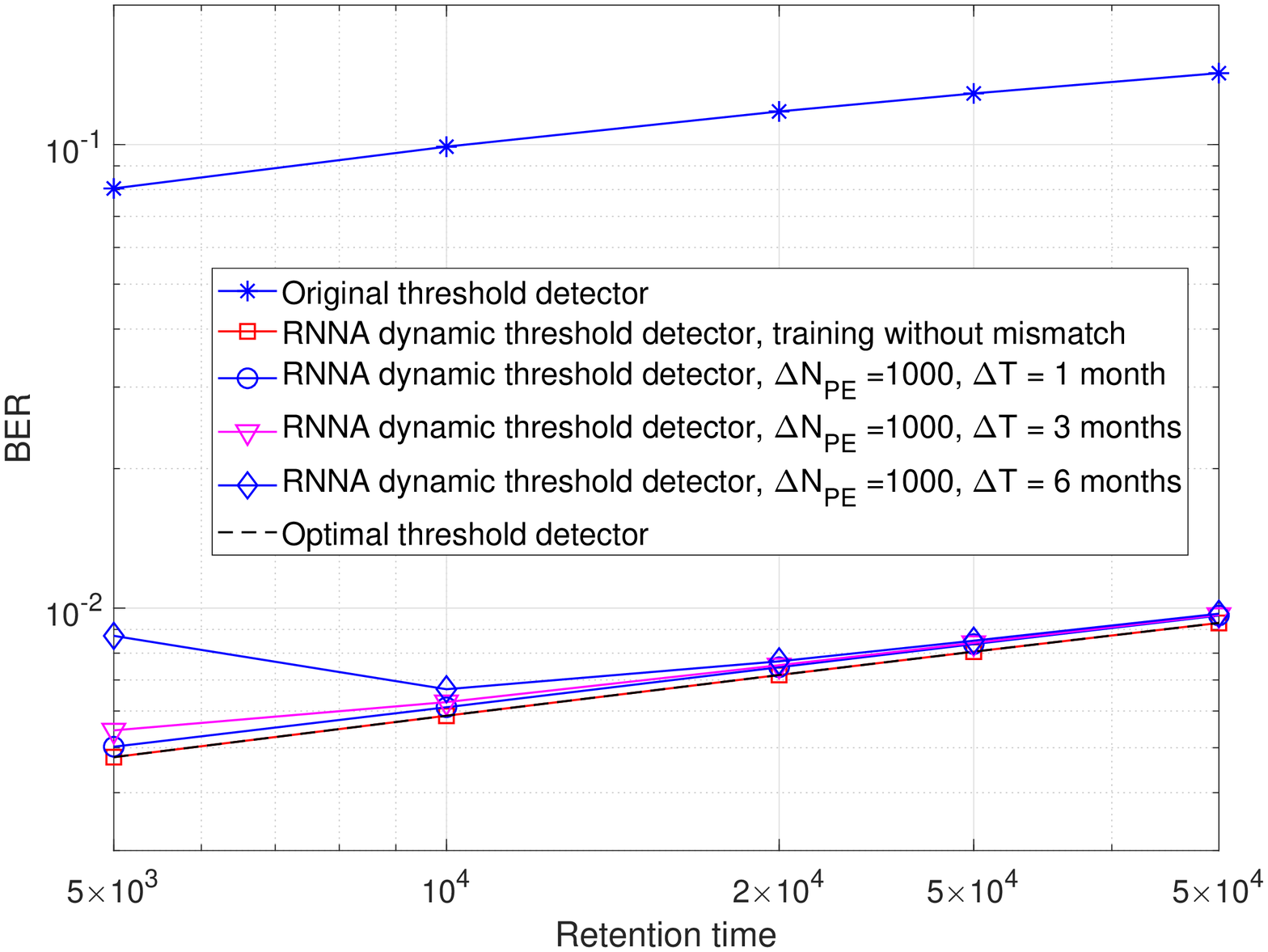}
\caption{BER comparison of different detectors. The RNNA dynamic threshold detector is trained with both P/E cycles and retention time mismatch, with $N^{\text{test}}_{\text{PE}}=10^{4}$.}
\label{ser_mismatch_pe_time}
\end{figure}

\section{PERFORMANCE EVALUATIONS}

In this work, the implementation and training of the RNN are performed using the machine learning library Keras \cite{keras}, with TensorFlow \cite{tensor} as the back-end. As described in Section IV.A, to reduce the read latency and power consumption, the RNN is trained periodically when the system is in the idle state. However, there might be a mismatch of P/E cycles and/or the retention time between the training set and test set. To measure the performance of the RNN detector and RNNA dynamic threshold detector in the presence of training and testing mismatch, we define the mismatch of P/E cycles as $\bigtriangleup  N_{\text{PE}}= N_{\text{PE}}^{\text{test}} - N_{\text{PE}}^{\text{train}}$, where $N_{\text{PE}}^{\text{train}}$ and $N_{\text{PE}}^{\text{test}}$ are the numbers of P/E cycles during training and testing, respectively. Similarly, we define the mismatch of the retention time as $\bigtriangleup  T= T^{\text{test}} - T^{\text{train}}$. In the following subsections, the uncoded/raw channel BER and LDPC-coded BER for MLC flash memory channels with our proposed RNNA quantizers are investigated.

\begin{figure}[t] 
\centering
\includegraphics[height=2.3in,width=3.6in]{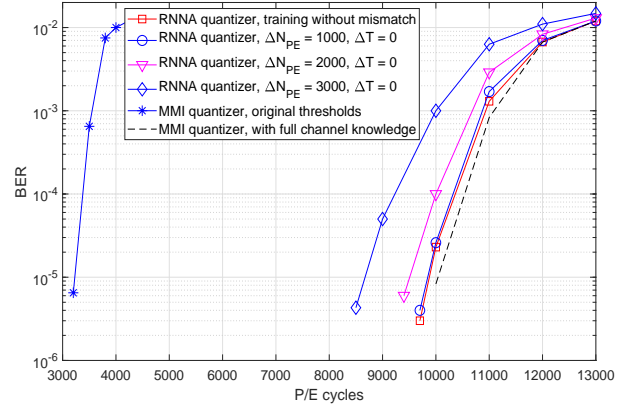}
\caption{BERs of LDPC codes with the RNNA quantizer, trained with different P/E cycles mismatch $(\bigtriangleup N_{\text{PE}}=1000, 2000, 3000)$,  at $ T^{\text{test}}=1\times 10^{4}$ hours.}
\label{ldpc_mismatch_pe}
\end{figure}

\subsection{Uncoded BER Performance}

First, to evaluate the effect of error corrupted labels on the performance of the RNN detector, we consider two cases: training with correct labels and training with error corrupted labels. As shown by Fig. \ref{error_label}, with a high LDPC decoding error rate of $5\times 10^{-3}$, the RNN detector can still approach the performance of the RNN detector trained with correct labels. This indicates that the RNN detector is robust to erroneous labels. 

Fig. \ref{ser_d} shows the BER performance of MLC flash memory channel using the RNN detector and RNNA dynamic threshold detector with thresholds obtained using (11). The optimum threshold detector we derived in Section IV.A is also included as a reference. Observe that the BER performance with the RNN detector is much better than that with the detector using the original thresholds (for $N_{\text{PE}}=0, T = 0$), and can closely approach the BERs of the optimum threshold detector. Furthermore, the proposed RNNA dynamic threshold detector even slightly outperforms the RNN detector and almost achieves the the performance of the optimum detector. Therefore, in the following simulations, we only employ the RNNA dynamic threshold detector, instead of the RNN detector.

Fig. \ref{ser_mismatch_pe} depicts the BERs of the RNNA dynamic threshold detector, where the RNN is trained with different amount of mismatch of P/E cycles between the training set and test set. Observe that when the mismatch $\bigtriangleup  N_{\text{PE}}$ is less than $1000$, the BER performance of the RNNA dynamic threshold detector can still closely approach the optimum performance. As the mismatch further increases, performance gap between the RNNA dynamic threshold detector and the optimum threshold detector becomes larger. In particular, this gap becomes significant at low BER regions when $\bigtriangleup  N_{\text{PE}}=3000$. This indicates that we only need to activate the RNN detector as well as the search of the adjusted read threshlds when the P/E cycle mismatch is greater than 3000.

To investigate the performance of the RNNA dynamic threshold detector in the presence of different retention time mismatch, we fix the mismatch of P/E cycles at 1000 and vary the the mismatch of retention time. As illustrated by Fig. \ref{ser_mismatch_pe_time}, when the retention time mismatch is 1 month, the RNNA dynamic threshold detector approaches the performance of the optimum detector. When the retention time mismatch increases to 3 months, the RNNA dynamic threshold detector only shows slight performance degradation. However, as the mismatch further increases to 6 months, a severe error floor occurs at $T=5\times 10^{3}$ hours. This indicates that we only need to activate the RNN detector as well as the search of the adjusted read thresholds every 3 months to maintain the near optimum BER performance.

\begin{figure}[t] 
\centering
\includegraphics[height=2.3in,width=3.6in]{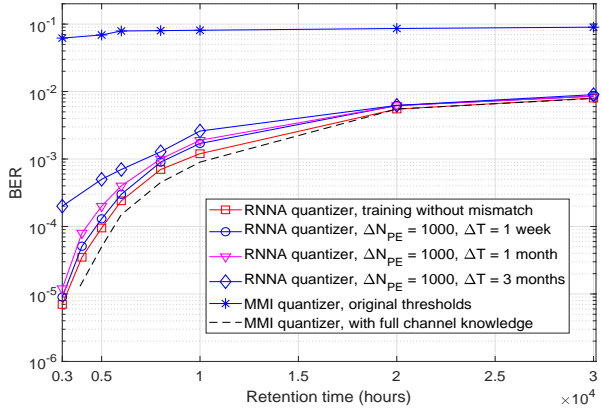}
\caption{BERs of LDPC codes with the RNNA quantizer, trained with both P/E cycles and retention time mismatch $(\bigtriangleup N_{\text{PE}}=1000, \bigtriangleup T=1$ week, 1 month and 3 months), and  at $N_{\text{PE}}^{\text{test}}=1.1\times 10^{4}$.}
\label{ldpc_mismatch_time}
\end{figure}

\subsection{LDPC-coded BER Performance}
In this subsection, we investigate the LDPC-coded performance using our proposed RNNA quantizer and the MMI quantizer. For cases with the MMI quantizer, we make the unrealistic assumption that the channel PDFs are known, and use the corresponding BERs as the performance benchmark. In the practical flash memories with the RNNA quantizer, the channel is unknown and the LLR is calculated using our proposed integer-based reliability mappings given in Table II. A rate 0.93 regular $(5, 69)$ LDPC code is employed with codeword length $N=8832$ bits and information length $K=8196$ bits, respectively. The LDPC codes is constructed by the progressive edge-growth (PEG) algorithm and decoded by using the NMS decoding algorithm with normalization factor $\alpha=0.5$. The maximum number of decoding iterations is 10.

Fig. \ref{ldpc_mismatch_pe} shows the BER comparison of LDPC codes with the RNNA quantizer, trained with difference P/E cycles mismatch, and at $ T^{\text{test}}=1\times 10^{4}$ hours. We first observe that the BER performance using the MMI quantizer with the original thresholds (for $N_{\text{PE}}=0, T = 0$) is much worse than that using the other quantizers. This implies that it is essential to update the read thresholds to maintain the error correction capability. When training without mismatch, the LDPC-coded performance with the RNNA quantizer approaches to that with the MMI quantizer designed using the full channel knowledge. For cases with the P/E cycle mismatch, the BERs with the RNNA quantizer is close to those with the MMI quantizer, up till to a P/E cycle mismatch of $\bigtriangleup N_{\text{PE}}=2000$.  This demonstrates the robustness of the LDPC coded system with the RNNA quantizer with respect to the training P/E cycle mismatch. However, when the mismatch continues to increases, more performance degradation occurs. In particular, when $\bigtriangleup N_{\text{PE}}$ increases to 3000, the LDPC-coded performance is about 1000 P/E cycles worse than the that without training mismatch. In addition, it is also observed that the LDPC coded performance shown by Fig. \ref{ldpc_mismatch_pe} is consistent with the threshold detectors' performance illustrated by Fig. \ref{ser_mismatch_pe}.

Next, we further vary the retention time and consider the case that both the mismatch of P/E cycles and retention time between the training set and test set exists in the LDPC coded system. Observe from Fig. \ref{ldpc_mismatch_time} that the BER performance of LDPC codes with the RNNA quantizer can also approach that with the MMI quantizer over a wide range of retention time. For the RNNA quantizer, with a P/E cycle mismatch of $\bigtriangleup N_{\text{PE}}=1000$, the LDPC code's performance degradation is small. However, when the retention time mismatch increases to 3 months, the BER performance is significantly degraded when the retention time is less than $5000$ hours. Again, we observe that the LDPC coded performance shown by Fig. \ref{ldpc_mismatch_time} is consistent with the threshold detectors' performance illustrated by Fig. \ref{ser_mismatch_pe_time}. 


\section{Conclusions}
We have considered to tackle the various non-stationary noises and the unknown offset of the NAND flash memory channels by using the DL techniques. In particular, we have first proposed a novel RNN-based detector, which can effectively detect the data symbols stored in the memory cell without any prior knowledge of the channel. However, compared with the conventional threshold detector, the RNN detector will lead to much longer read latency and more power consumption. To tackle this problem, we have proposed an RNNA dynamic threshold detector, whose detection thresholds can be derived based on the outputs of the RNN detector. In this way, we only need to activate the RNN detector periodically when the system is in the idle state. To enable SDD of ECCs, more read thresholds are required to generate the LLRs of the channel. In this work, we have first shown how to obtain more read thresholds based on the hard-decision read thresholds derived from the RNN detector. We then proposed integer-based reliability mappings based on the designed read thresholds, which can generate LLRs of the flash memory channel. Finally, to optimize the read thresholds in terms of the decoding performance, we have proposed to apply DE combined with differential evolution algorithm for the LDPC coded flash memory channels. Simulation results have shown that the BER performance of our proposed RNNA dynamic threshold detector can almost achieve that of the optimum threshold detector designed with the full knowledge of the channel. The BERs of the LDPC-coded system with the proposed RNNA dynamic read thresholds can closely approach that of the MMI quantizer which requires the full knowledge of the channel.
 We have also found that the proposed RNNA dynamic read thresholds design is robust to the training set and test set mismatch. Moreover, it only needs to be activated every few months. Hence it shows a high potential for practical applications.

\small
\bibliographystyle{IEEEtran}
\bibliography{postdoc_refs}

\end{document}